\documentclass[a4paper]{article}

\usepackage[utf8]{inputenc}
\usepackage[russian]{babel}
\usepackage{amssymb, amsfonts, amsthm, amsmath}
\usepackage{graphicx}

\begin{document}

\title{Волновая функция фотоэлектрона вблизи центра квантового вихря}

\author{Н.В. Ларионов, Ю.Л. Колесников}

\abstract{В двумерном приближении теоретически исследуются плотность и ток вероятности для фотоэлектрона в области локализации квантового вихря. Волновая функция в импульсном представлении, найденная нами ранее, упрощается вблизи нуля, соответствующего  центру вихря. Это позволяет получить для неё простое аналитическое выражение, имеющее структуру гауссова волнового пакета и содержащее основную информацию о вихре. С его помощью анализируется временная эволюция квантового вихря, как в импульсном, так и координатном пространстве. Также исследуется влияние величины напряжённости ионизирующего сверхкороткого лазерного импульса на геометрию квантового вихря.}

\maketitle

\section{Введение}

Ранее нами исследовались квантовые вихри, образующиеся при надбарьерной ионизации двумерного атома водорода сверхкоротким лазерным импульсом (см. \cite{1}-\cite{5} и ссылки там же). Расчёты проводились как с помощью численного решения уравнения Шредингера, так и с помощью аналитического выражения для волновой функции фотоэлектрона, полученной в рамках второго порядка нестационарной теории возмущений. Эта волновая функция изначально выведена в импульсном $k$-представлении и идентификация центров квантовых вихрей, а также анализ "<симметричного"> потока вероятности \cite{6}, проводились в соответствующем $k$-пространстве.
Переход из $k$-пространства в обычное координатное пространство, для упомянутой волновой функции, выполнен не был. Причина этому -- существенные трудности при двумерном преобразовании Фурье.

В данной статье, для случая хорошо локализованных вихрей, нам удаётся упростить упомянутую волновую функцию фотоэлектрона таким образом, что она записывается в виде произведения гауссовой функции от модуля импульса  $k=|\mathbf{k}|$ на полином второй степени относительно проекций $k_{x}$, $k_{y}$. Такая упрощённая волновая функция не теряет информацию о квантовом вихре и позволяет легко исследовать его и в координатном пространстве.

В работе традиционно используется атомная система единиц: $\hbar=1$, $m_{e}=1$, $e=1$.

\section{Теоретическая модель }

Рассматриваемая нами модель представлена двумерным атомом водорода \cite{7}, взаимодействующим со сврехкоротким лазерным импульсом, напряжённость $\mathbf{F}(t)$ электрического поля которого равна
\begin{gather}
   \mathbf{F}(t)=\left(F_{x}(t),0\right)=
    \nonumber \\
   =\mathbf{e}_{x}F_{0}\cos(\omega t -\alpha )
   \left[\theta(T-t) - \theta(-t)\right], \label{Laser}
\end{gather}%
где $\mathbf{e}_{x}$  –  единичный вектор в направлении оси $x$, $\omega$ –  частота,  $\theta(t)$  –  функция Хевисайда,   $T$ – длительность импульса,  $\alpha$  –  начальная фаза. Значения постоянной амплитуды $F_{0}$ берутся такими чтобы преобладает надбарьерная ионизация.

Волновая функция фотоэлектрона в  $k$-представлении, полученная нами ранее в рамках второго порядка нестационарной теории возмущений \cite{4,5}, имеет вид
\begin{gather}
    \tilde{\Psi}(\mathbf{k},t)= \sum_{m} \int_{0}^{\infty}b_{k',m}(t)\Psi_{k',m}^{(0)}(\mathbf{k}) e^{-i E_{k'}t}k'dk'=
     \nonumber \\
    -i\sqrt{\frac{2}{\pi}}b_{k,1,10}^{(1)}(t)\cos(\varphi_{k})e^{-i E_{k}t}+
    \nonumber \\
   + \frac{1}{\sqrt{2\pi}}b_{k,0,10}^{(2)}(t)e^{-i E_{k}t} -
    \label{Psi} \\
    -\sqrt{\frac{2}{\pi}}b_{k,2,10}^{(2)}(t)\cos(2\varphi_{k})e^{-i E_{k}t}. \nonumber
\end{gather}%
Здесь $(k,\varphi_{k})$  - полярные координаты импульса $\mathbf{k}$ фотоэлектрона с энергией
\begin{equation}
   E_{k}=k^{2}/2=(k_{x}^{2}+k_{y}^{2})/2, \nonumber
\end{equation}%
и проекцией момента на ось ось $z$ $m=0,\pm1,\pm2,...$.

$\Psi_{k',m}^{(0)}(\mathbf{k})$ - цилиндрическая волна с амплитудой
\begin{equation}
   b_{k',m}(t)=b_{k',m,10}^{(1)}(t)+b_{k',m,10}^{(2)}(t), \nonumber
\end{equation}%
где верхние индексы $i=1,2$ указывают на порядок теории возмущений, а нижний индекс «$10$» указывает на начальное связанное состояние электрона. Отметим, что в (\ref{Psi}) полностью игнорируется связь с атомным остатком.

Амплитуды $b_{k,m,10}^{(i)}(t)$ могут быть записаны в следующем виде
\begin{equation}
    b_{k,1,10}^{(1)}(t) =
   \frac{-3ik}{(k^{2}+1)^{5/2}}\int_{0}^{t}dt'F_{x}(t')e^{i \omega_{k1}t'}, \label{b1}
\end{equation}%

\begin{gather}
   b_{k,0,10}^{(2)}(t) = (-i)\int_{0}^{t}dt'F_{x}(t')\times
   \nonumber\\
   \times\left(\frac{\partial}{\partial k}-ik t'+\frac{1}{k}\right)b_{k,1,10}^{(1)}(t'),
   \nonumber \\
   b_{k,2,10}^{(2)}(t) =\frac{i}{2}\int_{0}^{t}dt'F_{x}(t')\times
   \label{b2} \\
   \times\left(\frac{\partial}{\partial k}-ik t'-\frac{1}{k}\right)b_{k,1,10}^{(1)}(t') +
    \nonumber
\end{gather}%
где $\omega_{k1}=(k^{2}+1)/2$ -- частота перехода из связанного состояния в непрерывный спектр.

Рассмотрим случай установившегося решения $t>T$ и выберем такие параметры $T$, $w$, $\alpha$ лазерного импульса, для которых ранее были идентифицированы хорошо локализованные  квантовые вихри \cite{1,5}.

Так для $T=4$, $w=\pi$, $\alpha=0$ интегралы (\ref{b1}), (\ref{b2}) легко берутся и волновая функция фотоэлектрона (\ref{Psi}) принимает вид
\begin{gather}
   \tilde{\Psi}(k,\varphi_{k},t)= A \cdot \frac{\sin(k^{2}+1)}{(k^{2}+1)^{3/2}}
   e^{i k^{2} -i E_{k}t}\times
    \nonumber \\
    \times\left[\frac{k\cos(\varphi_{k})}{(k^{2}+1)^{2}-4\pi^{2}}
    \left(1+\frac{2 i F_{0}k\cos(\varphi_{k})(7(k^{2}+1)^{2}-4\pi^{2})}{(k^{2}+1)^{2}((k^{2}+1)^{2}-16\pi^{2})}\right)-
    \right. \nonumber \\
    \left.- \frac{2 i F_{0}}{(k^{2}+1)((k^{2}+1)^{2}-16\pi^{2})}\right],
    \label{PsiT4Wpi}
\end{gather}%
где $A$ - константа. Здесь члены с $F_{0}$ соответствуют второму порядку теории возмущений, а члены свободные от этой амплитуды -- первому порядку.

Для используемых параметров импульса центры двух симметричных квантовых вихрей, являющиеся нулями волновой функции (\ref{PsiT4Wpi}), имеют декартовы координаты $k_{x_{0}}= 0, k_{y_{0}}= \pm \sqrt{2\pi-1}$  или следующие полярные координаты  $k_{0}= \sqrt{2\pi-1}, \varphi_{0}= \pi/2, 3\pi/2$. А векторное поле "<симметричного"> потока
\begin{equation}
   \overline{\mathbf{j}}(\mathbf{k},t)=\textrm{Im}[\tilde{\Psi}^{*}(\mathbf{k},t) \nabla_{k}\tilde{\Psi}(\mathbf{k},t)],
   \label{SymFlux}
\end{equation}%
где $\nabla_{k}\equiv\partial/\partial \mathbf{k}$, закручивается вокруг оси $z$, проходящей через эти центры.

Также можно легко убедиться, что в рассматриваемом случае средний импульс фотоэлектрона в состоянии (\ref{PsiT4Wpi}) равен нулю
\begin{equation}
   \langle k_{x,y}\rangle=\int k_{x,y}|\tilde{\Psi}(\mathbf{k},t)|^{2} \frac{d^{2}k}{2\pi}= 0,
   \label{AvPulse}
\end{equation}%
где $k_{x}=k\cos(\varphi_{k})$,  $k_{y}=k\sin(\varphi_{k})$.

Это результат легко понять если вспомнить взятые параметры лазерного импульса и приближения используемые при выводе волновой функции фотоэлектрона (\ref{Psi}). С математической точки зрения выражение (\ref{AvPulse}) содержит следующие интегралы
\begin{gather}
   \int_{0}^{2\pi} \cos(\varphi_{k})\cos(n \varphi_{k}) d\varphi_{k},
   \nonumber \\
   \int_{0}^{2\pi} \sin(\varphi_{k})\cos(n \varphi_{k}) d\varphi_{k},
   \nonumber
\end{gather}%
которые в нашем случае $n=0,2,4$, обращают его в ноль.

Дисперсия любой из компонент импульса не равна нулю
\begin{equation}
   \langle k_{x,y}^{2}\rangle=\int k_{x,y}^{2}|\tilde{\Psi}(\mathbf{k},t)|^{2} \frac{d^{2}k}{2\pi}\neq 0.
   \label{AvPulse2}
\end{equation}%

Интеграл (\ref{AvPulse2}) не удаётся вычислить аналитически. Ниже сравним значения численного решения (\ref{AvPulse2}) с приближённым решениями.

Несмотря на относительно простой вид волновой функции (\ref{PsiT4Wpi}) при попытке записать её в координатном представлении наталкиваешься на существенные трудности при преобразовании Ханкеля \cite{8}. Именно по этой причине исследование вихрей в обычном пространстве проводилось нами ранее только с помощью численного решения уравнения Шредингера \cite{1}.

Далее будем интересоваться поведением фотоэлектрона вблизи центров квантовых вихрей. Для этого разложим волновую функцию (\ref{PsiT4Wpi}) в ряд Тейлора вблизи $k_{0}$. Так для отдельных функций  из (\ref{PsiT4Wpi}) имеем
\begin{gather}
   \sin(k^{2}+1)\approx (k^{2}-k_{0}^{2}),
    \nonumber \\
    \frac{1}{(k^{2}+1)^{3/2}(k^{2}+(2\pi+1))}=
    \nonumber \\
    =e^{-\ln\left[(k^{2}+1)^{3/2}(k^{2}+(2\pi+1))\right]}\approx
    \nonumber \\
    \approx \frac{1}{8\sqrt{2}\pi^{5/2}}e^{-\frac{k^{2}-k_{0}^{2}}{\pi}},
    \nonumber \\
    \frac{1}{(k^{2}+1)^{5/2}((k^{2}+1)^{2}-16\pi^{2})}=
    \nonumber \\
    =e^{-\ln\left[ (k^{2}+1)^{5/2}((k^{2}+1)^{2}-16\pi^{2})\right]}\approx
    \nonumber \\
    \approx  -\frac{1}{48\sqrt{2}\pi^{9/2}}e^{-\frac{k^{2}-k_{0}^{2}}{\pi}}.
    \label{SerTaylor}
\end{gather}%
Учтём, что $((k^{2}+1)^{2}-4\pi^{2})=(k^{2}-k_{0}^{2})(k^{2}+(2\pi+1))$ и опустим малое слагаемое пропорциональное $\cos^{2}(\varphi_{k})$.

Тогда волновая функция фотоэлектрона (\ref{PsiT4Wpi}) вблизи центра вихря, примет следующий вид
\begin{gather}
   \tilde{\Psi}(k,\varphi_{k},t)= A \cdot
   e^{-\frac{k^{2}-k_{0}^{2}}{\pi}+i k^{2}-i E_{k}t}\times
    \nonumber \\
    \times\left[  k\cos(\varphi_{k}) +
    \frac{i F_{0} (k^{2}-k_{0}^{2})}{3\pi^{2}}\right].
    \label{SerPsik0}
\end{gather}%

Полученное выражение (\ref{SerPsik0}) представляет собой гауссову функцию умноженную на полином второй степени относительно компонент импульса. Как видно, этот полином определяет нули волновой функции, соответствующие центрам вихрей $(k_{0}, \varphi_{0})$. В самом деле, для нахождения значений $(k_{0}, \varphi_{0})$ достаточно приравнять мнимую и действительную части полинома нулю \cite{Ovch2010}-\cite{Ovch2014}.

Вычисляя с помощью (\ref{SerPsik0}) средний импульс фотоэлектрона убеждаемся, что как и в случае с точной волной функцией (\ref{PsiT4Wpi}), он равен нулю $\langle k_{x,y}\rangle=0$. Для дисперсии получаем следующие выражения
\begin{gather}
   \langle k_{x}^{2}\rangle=\frac{\pi}{4}
   \frac{\left[27\pi^{5} +F_{0}^{2}(4-8\pi+6\pi^{2})\right]}{9\pi^{5} +2F_{0}^{2}(2-6\pi+5\pi^{2})}\approx \frac{3\pi}{4},
   \nonumber \\
   \langle k_{y}^{2}\rangle=\frac{\pi}{4}
   \frac{\left[9\pi^{5} +F_{0}^{2}(4-8\pi+6\pi^{2})\right]}{9\pi^{5} +2F_{0}^{2}(2-6\pi+5\pi^{2})}\approx \frac{\pi}{4}.
   \label{SerAvPP2}
\end{gather}%

Видно, что зависимость от напряжённости поля очень слабая. Если сравнить приближённые значения дисперсии (\ref{SerAvPP2}) с соответствующими численными значениями, полученными с помощью (\ref{PsiT4Wpi}), то последние примерно в $1.2$ раза больше приближённых.

Основное преимущество выражения (\ref{SerPsik0}), по сравнению с его точным аналогом (\ref{PsiT4Wpi}), заключается в том, что его легко можно переписать в координатном представлении. Не вдаваясь в подробности простых вычислений выпишем ответ
\begin{gather}
   \psi(r,\varphi,\tau)=
   \nonumber \\
   \int_{0}^{\infty}\int_{0}^{2\pi}\tilde{\Psi}(k,\varphi_{k},\tau)
   e^{i kr \cos(\varphi_{k}-\varphi)}\frac{kdkd\varphi_{k}}{(2\pi)^{2}} =
   \nonumber \\
   = \frac{\tilde{A}}{a^{3}(\tau)} \cdot e^{-\frac{r^{2}}{a^{2}(\tau)}
   +i \frac{\pi r^{2}\tau}{2a^{2}(\tau)}}\times
    \nonumber \\
    \times\left[ \left(F_{0}(4(\pi-1)+\pi^{2}(r^{2}-k_{0}^{2}\tau^{2}))-6\pi^{3}r\cos(\varphi)\right)
    +
    \right.
    \nonumber \\
    \left.
    + i \pi \tau \left(2F_{0}(3\pi-2)-3\pi^{3}r\cos(\varphi)\right)\right],
    \label{SerPsiRk0}
\end{gather}%
где $\tilde{A}$ - константа, $\mathbf{r}=(r,\varphi)$  - полярные координаты фотоэлектрона и введены обозначения:  $a^{2}(\tau)=(4+\pi^{2}\tau^{2})/\pi$, $\tau=t-2$.

Используя (\ref{SerPsiRk0}) легко находим средние значения координат фотоэлектрона и их дисперсии
\begin{gather}
   \langle x\rangle \approx -\frac{12k_{0}^{2}}{9\pi^{5}}F_{0},\langle y\rangle=0
    \nonumber \\
  \langle (\triangle x)^{2}\rangle =\langle x^{2}\rangle-\langle x\rangle^{2}\approx \frac{3\pi\tau^{2}}{4},
  \nonumber \\
  \langle (\triangle y)^{2}\rangle =\langle y^{2}\rangle-\langle y\rangle^{2}\approx \frac{\pi\tau^{2}}{4}.
   \label{SerAvXX2}
\end{gather}%
При этом, как и должно быть, $a^{2}(\tau)\approx \langle (\triangle x)^{2}\rangle + \langle (\triangle y)^{2}\rangle$.

Полученная волновая функция в координатном представлении (\ref{SerPsiRk0}) имеет сходную структуру с волновой функцией в импульсном представлении (\ref{SerPsik0}). Гауссов множитель описывает расплывание образованного волнового пакета. Полином второй степени, относительно координат фотоэлектрона, несет информацию о квантовом вихре в обычном пространстве. Приравнивая нулю реальную и мнимую части выражения в квадратных скобках в (\ref{SerPsiRk0}) и решая получившуюся систему уравнений, находим координаты центров вихрей
\begin{equation}
  x_{0}=\frac{F_{0}(6\pi-4)}{3\pi^{3}},y_{0}\approx \pm k_{0}\tau .
    \label{VortexCenterR}
\end{equation}%

В следующем разделе с помощью полученной волновой функции (\ref{SerPsiRk0}) исследуем структуру вихрей в координатном пространстве.

\section{Результаты расчётов}

\textit{Импульсное пространство.} Вначале проверим, как изменилась плотность вероятности при разложении волновой функции вблизи центра вихря. Напомним, что везде момент времени $t$ берется таким, что $t>T=4$.

На рисунке 1 представлены графики плотности распределения фотоэлектрона по импульсам $\rho(\mathbf{k})=|\tilde{\Psi}(\mathbf{k},t)|^{2}$ (для более чёткого отображения графики строятся для $ \ln(\rho)$), построенной с помощью "<точной"> волновой функции (\ref{PsiT4Wpi}) Рис.1 a) и с помощью её приближённого выражения (\ref{SerPsik0}) Рис.1 b).
\begin{figure*}[th]
\includegraphics[width=15.0pc]{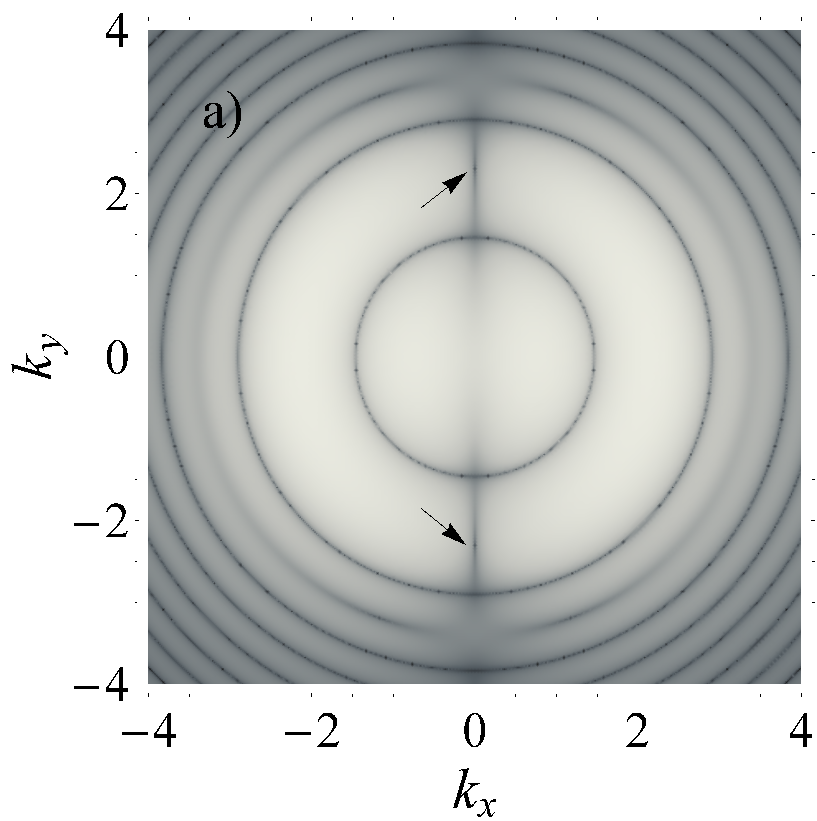}
\includegraphics[width=15.0pc]{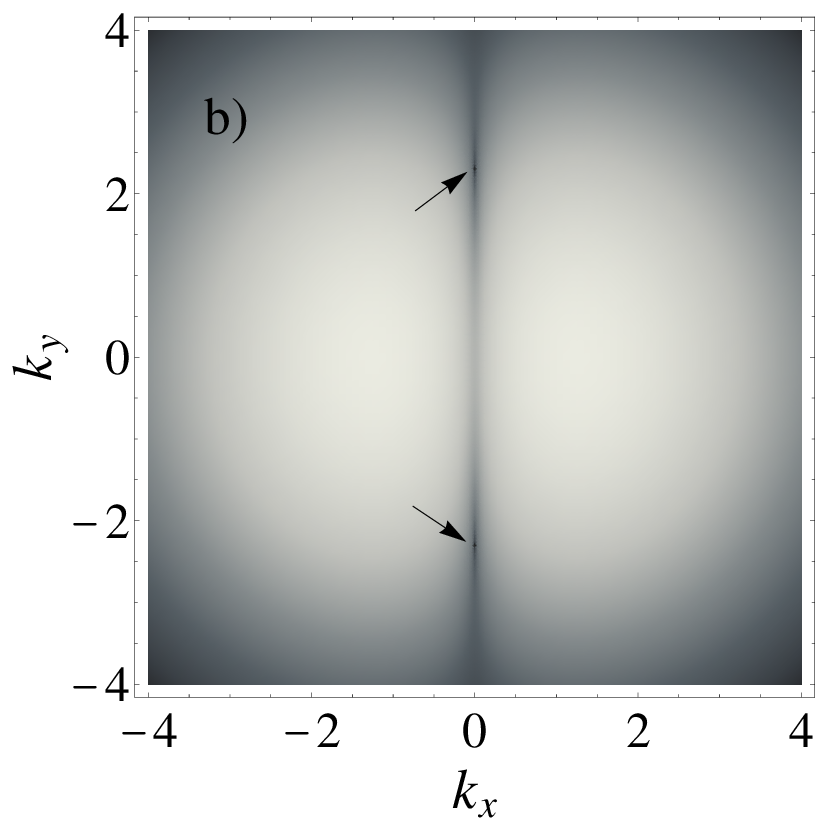}
\includegraphics[width=15.0pc]{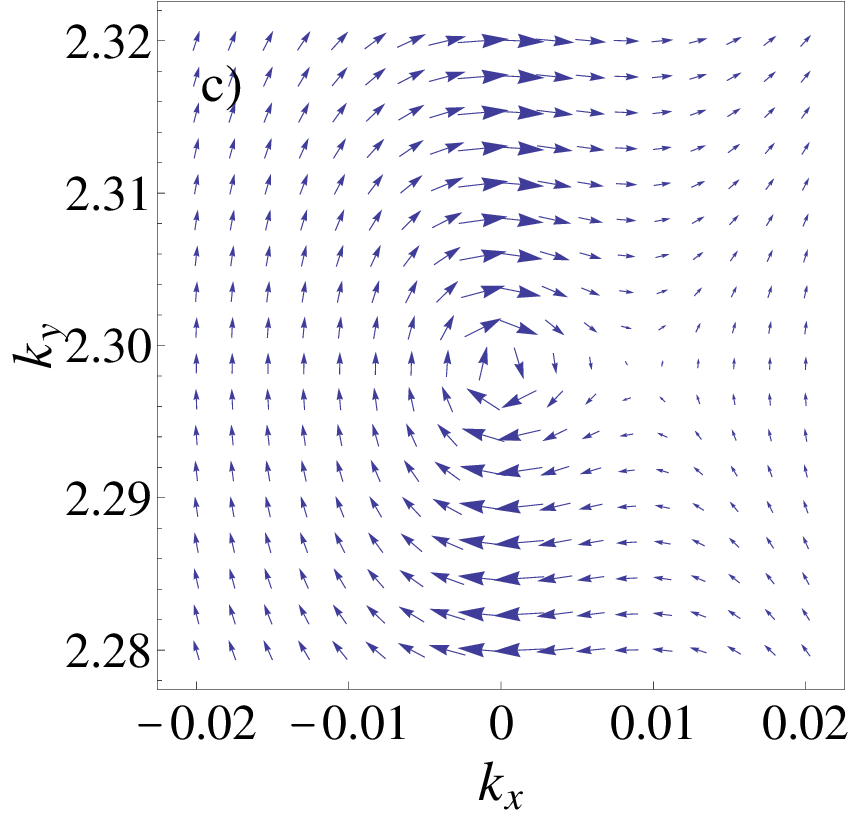}
\includegraphics[width=15.0pc]{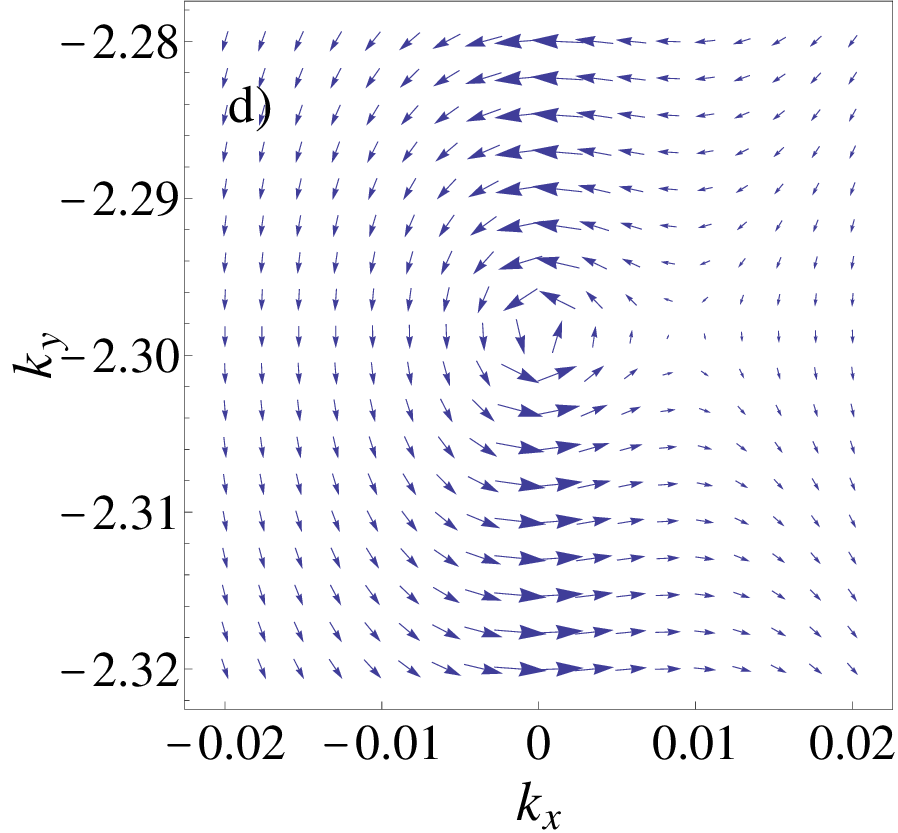}
\caption{\label{f1} a) и b) - плотность распределения фотоэлектрона по импульсам  $\ln[\rho(k_{x},k_{y})]$. c) и d) – векторное поле $\overline{\mathbf{v}}(k_{x},k_{y},t)$ вблизи центров квантовых вихрей. Напряжённость лазерного поля $F_{0}=0.4$. Момент времени $t=5$.}
\end{figure*}

Аппроксимация волновой функции (\ref{PsiT4Wpi}) гауссовой функцией привело к потере информации о состояниях фотоэлектрона, соответствующих цилиндрическим волнам. Однако информация о квантовых вихрях сохранилась, что видно не только по нулям плотности вероятности (стрелки), но и по характеру векторного поля Рис.1 c),d), построенного с помощью нормированного "<симметричного"> потока $\overline{\mathbf{v}}(\mathbf{k},t)=\overline{\mathbf{j}}(\mathbf{k},t)/\rho(\mathbf{k})$.

Здесь представлено поле $\overline{\mathbf{v}}(\mathbf{k},t)$ полученное только с помощью приближённой функции  (\ref{SerPsik0}). Для выбранной области значений $k_{x},k_{x}$ оно ничем не будет отличаться от поля, построенного с использованием "<точной"> функции (\ref{PsiT4Wpi}) (см. \cite{5}).

Заметим, что верхний и нижний вихри абсолютно одинаковы за исключением того, что направление вращение у них противоположны.

Также отметим, что в силу свободного движения фотоэлектрона плотность вероятности $\rho(\mathbf{k})$ записывается без аргумента $t$. Что же касается "<симметричного"> потока (\ref{SymFlux}), то в силу его чувствительности к фазе волновой функции \cite{6}, временная зависимость сохраняется.

\textit{Координатное пространство.} На следующем рисунке 2 представлены плотность вероятности распределения фотоэлектрона по координатам $\rho(\mathbf{r},t)=|\psi(\mathbf{r},t)|^{2}$ и поле скоростей фотоэлектрона
\begin{equation}
\mathbf{v}(\mathbf{r},t)=\textrm{Im}[\psi^{*}(\mathbf{r},t) \nabla \psi(\mathbf{r},t)]/\rho(\mathbf{r},t) \nonumber
\end{equation}
(термин "<поле скоростей"> заимствован из квантовой гидродинамики \cite{Madelung1926}-\cite{Madelung2009}), построенные с помощью найденной волновой функции фотоэлектрона в координатном представлении (\ref{SerPsiRk0}).
\begin{figure*}[th]
\includegraphics[width=15.0pc]{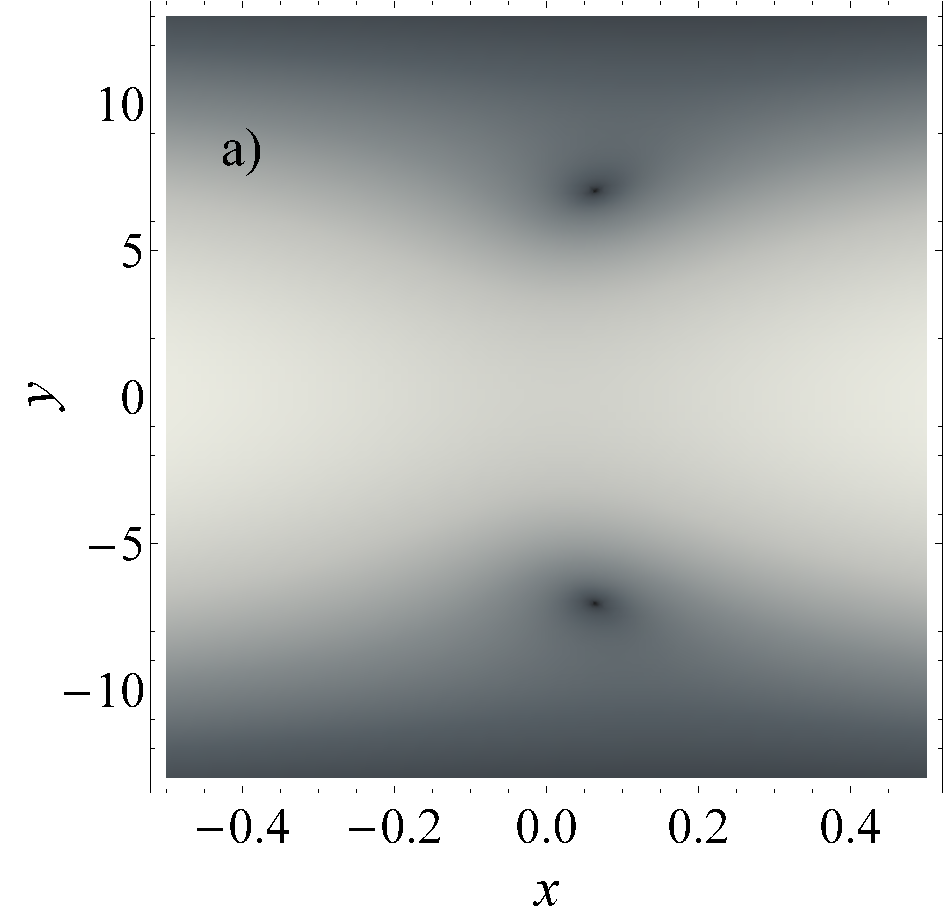}
\includegraphics[width=15.0pc]{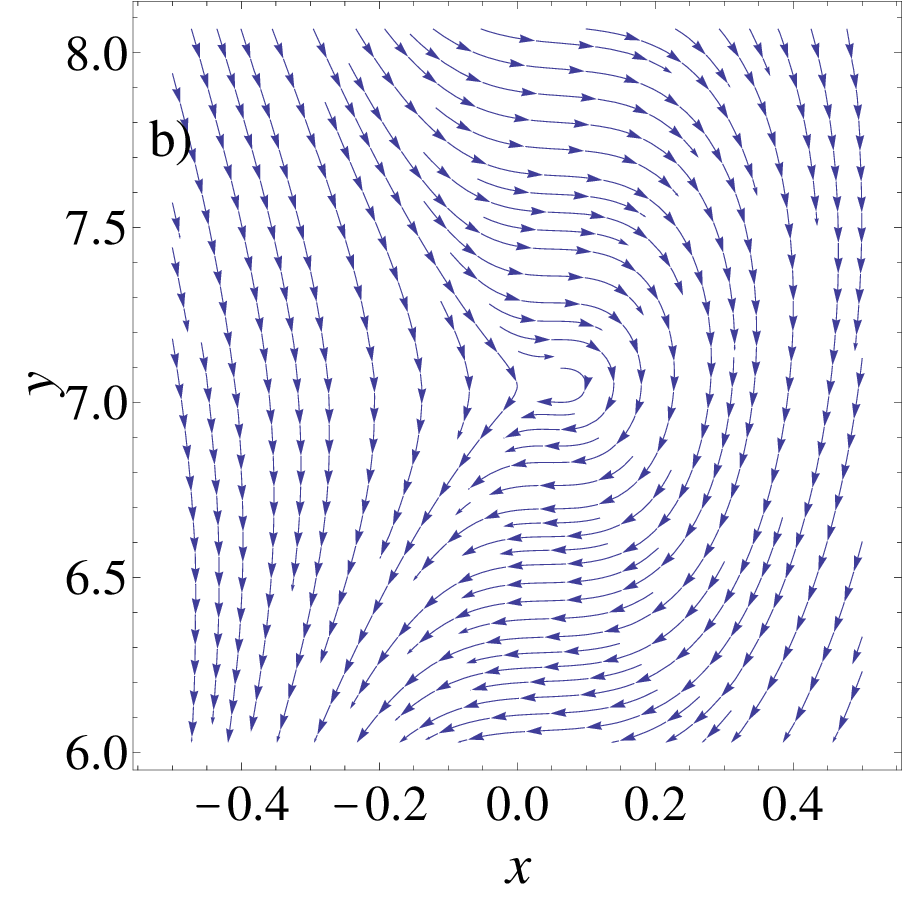}
\includegraphics[width=15.0pc]{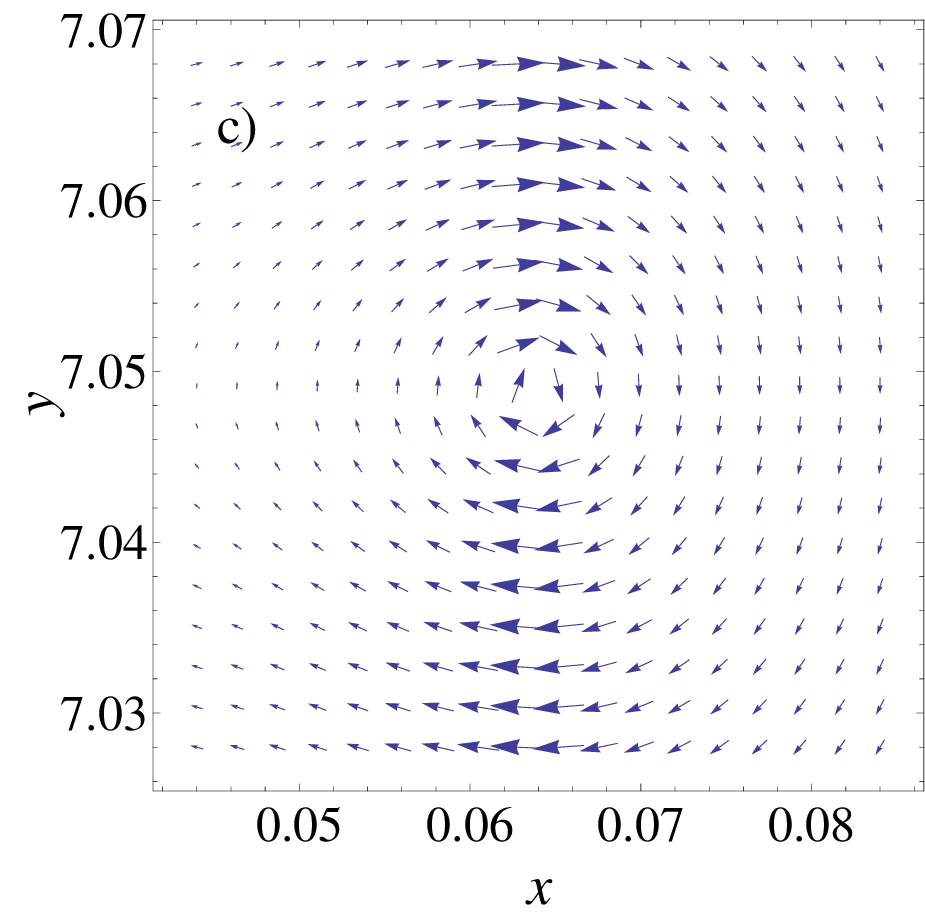}
\includegraphics[width=15.0pc]{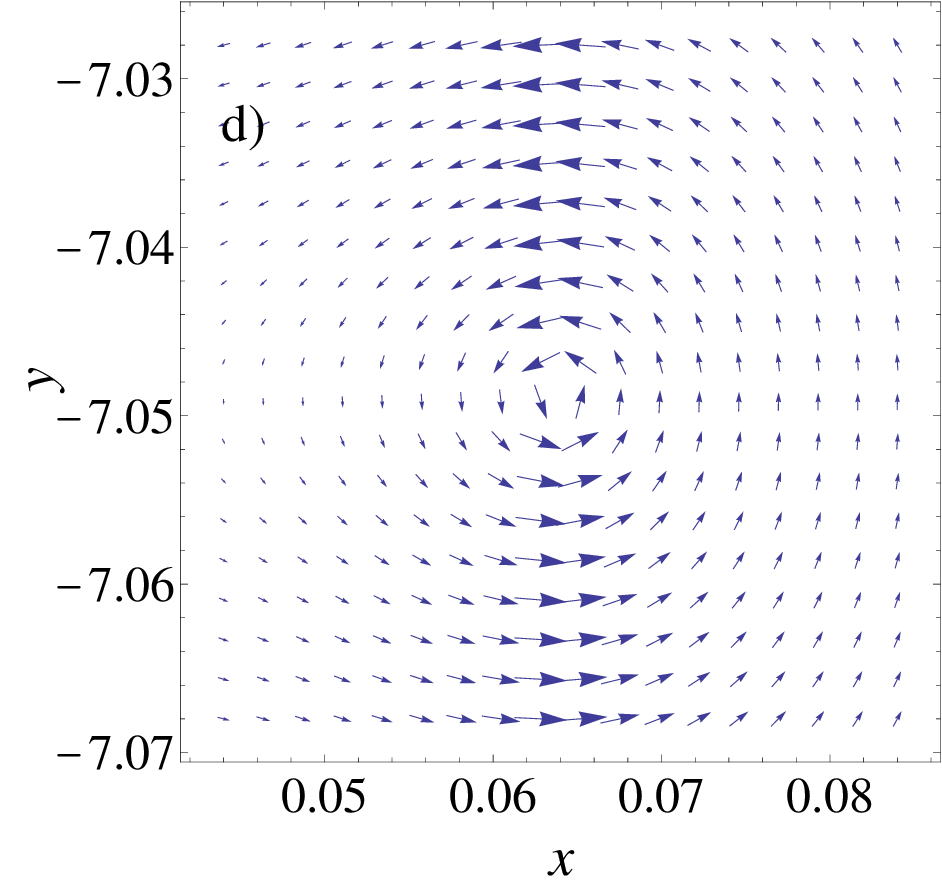}
\caption{\label{f2} a) - плотность распределения фотоэлектрона по координатам  $\ln[\rho(x,y,t)]$. b), c), d) – векторное поле $\mathbf{v}(x,y,t)$ вблизи центров квантовых вихрей. Напряжённость лазерного поля $F_{0}=0.4$. Момент времени $t=5$.}
\end{figure*}

Также как и в импульсном пространстве имеются два симметричных вихря, с противоположным направлением вращения Рис.2 c),d). Координаты центров вихрей даются формулами (\ref{VortexCenterR}) и равны $x_{0}=0.064$, $y_{0}=\pm 7.05$.

Отметим сходство структуры вихрей в импульсном и координатном пространствах. Однако реальная геометрия вихрей в координатном пространстве, для выбранных параметров импульса, более сложная \cite{1}. Предварительные расчёты показывают, что учёт слагаемого $\sim\cos^{2}(\varphi_{k})$, отброшенного нами при выводе (\ref{SerPsik0}), приведет к геометрии вихрей близкой к полученной нами ранее численно \cite{1}.

Теперь проследим временную эволюцию вихрей. На рисунке 3 для двух различных моментов времени $t_{1}=5$, $t_{2}=10$ представлено поле $\mathbf{v}(x,y,t_{i})$, а также следующие зависимости модуля волновой функции $b_{x}(t_{i})\equiv | \psi(x,y_{0},t_{i})|$, $b_{y}(t_{i})\equiv | \psi(x_{0},y,t_{i})|$ от одной из координат ($\tilde{b}$ - нормированный на свой максимум модуль волновой функции $b$).

Видно, что расплывание волнового пакета, отображенное на графиках Рис.3 c),d), не изменяет геометрию и масштаба вихря Рис.3 a),b). Вихрь перемещается в пространстве без изменений и координата его центра описывается формулой (\ref{VortexCenterR}): $t_{1}$ -- $x_{0}=0.064$, $y_{0}=\pm 7.05$; $t_{2}$ -- $x_{0}=0.064$, $y_{0}=\pm 18.446$.
\begin{figure*}[th]
\includegraphics[width=15.0pc]{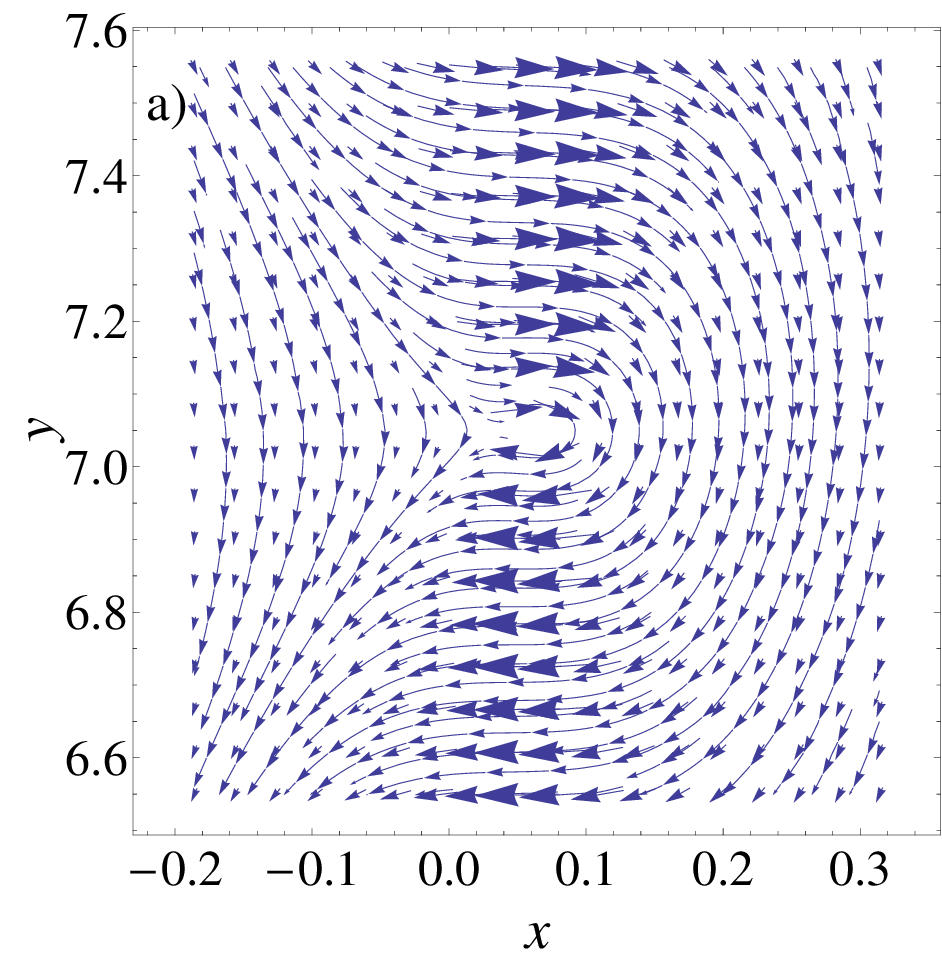}
\includegraphics[width=15.0pc]{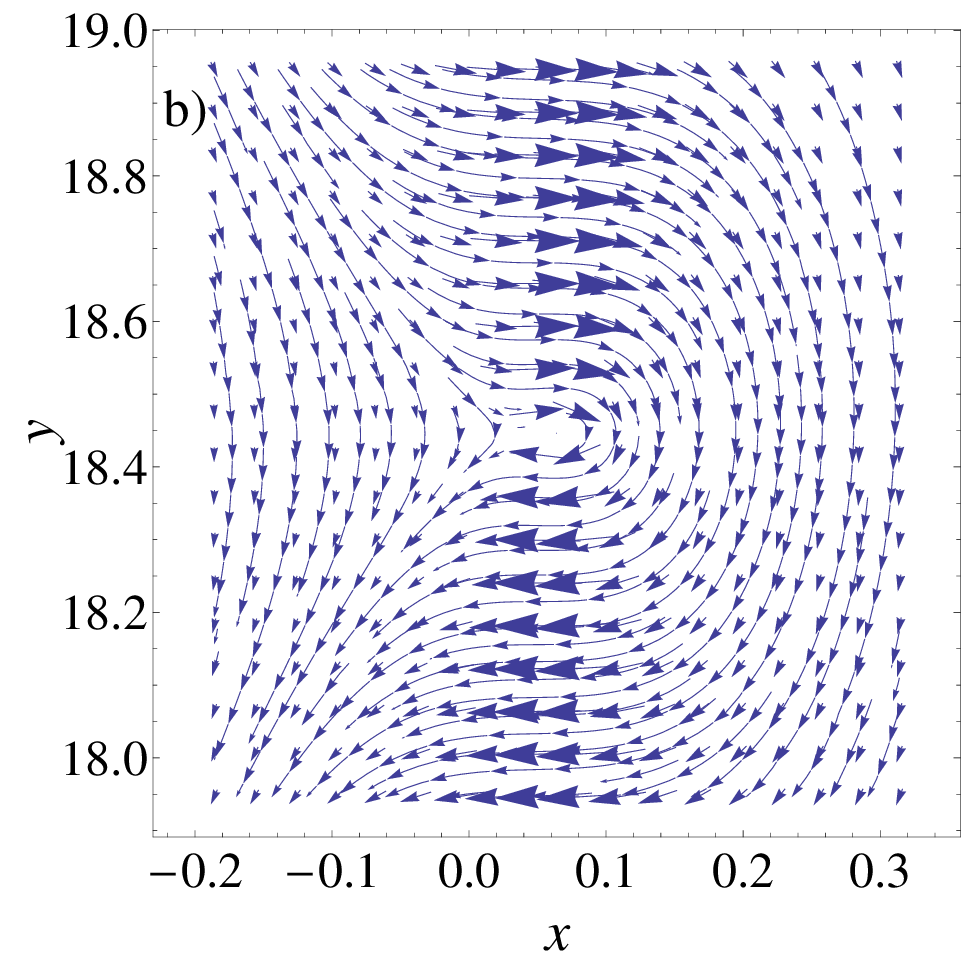}
\includegraphics[width=15.0pc]{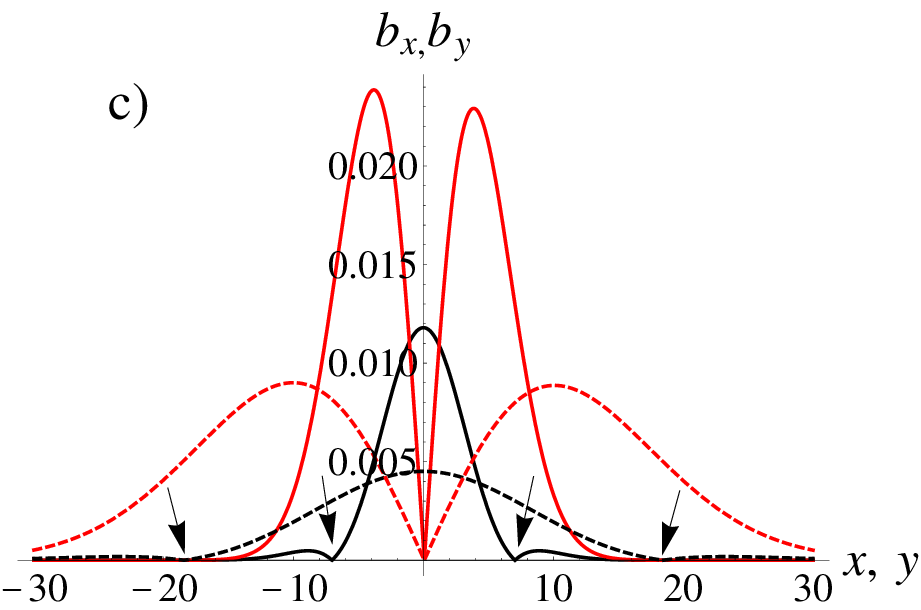}
\includegraphics[width=15.0pc]{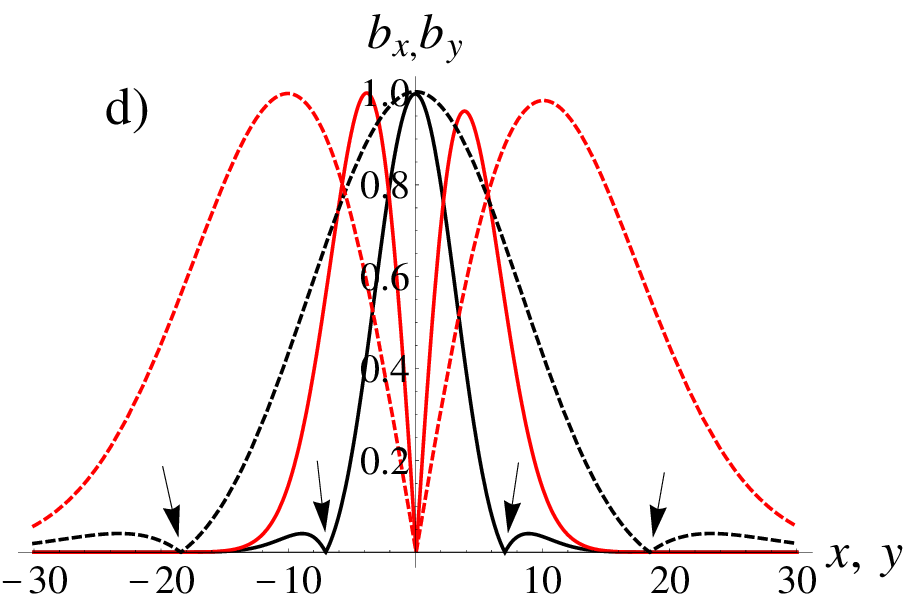}
\caption{\label{f3} Векторное поле $\mathbf{v}(x,y,t_{i})$ вблизи центров квантовых вихрей в разные моменты времени: a) $t_{1}=5$, b) $t_{2}=10$. c),d) -- зависимость абсолютных значений волновой функции $b_{x}$ (красная кривая), $b_{y}$ (чёрная кривая) от компонент $x$, $y$, соответственно: $t_{1}=5$ -- сплошная линия, $t_{2}=10$ -- пунктирная линия. Стрелками указаны центры вихрей. Напряжённость лазерного поля $F_{0}=0.4$. }
\end{figure*}

На следующем рисунке 4 построено поле скоростей фотоэлектрона $\mathbf{v}(x,y,t)$ при разных значениях напряжённости $F_{0}$ ионизирующего лазерного импульса. Здесь, учитывая внезапность включения поля (\ref{Laser}), мы выходим за пределы малых возмущений \cite{LandauV3}.
\begin{figure*}[th]
\includegraphics[width=15.0pc]{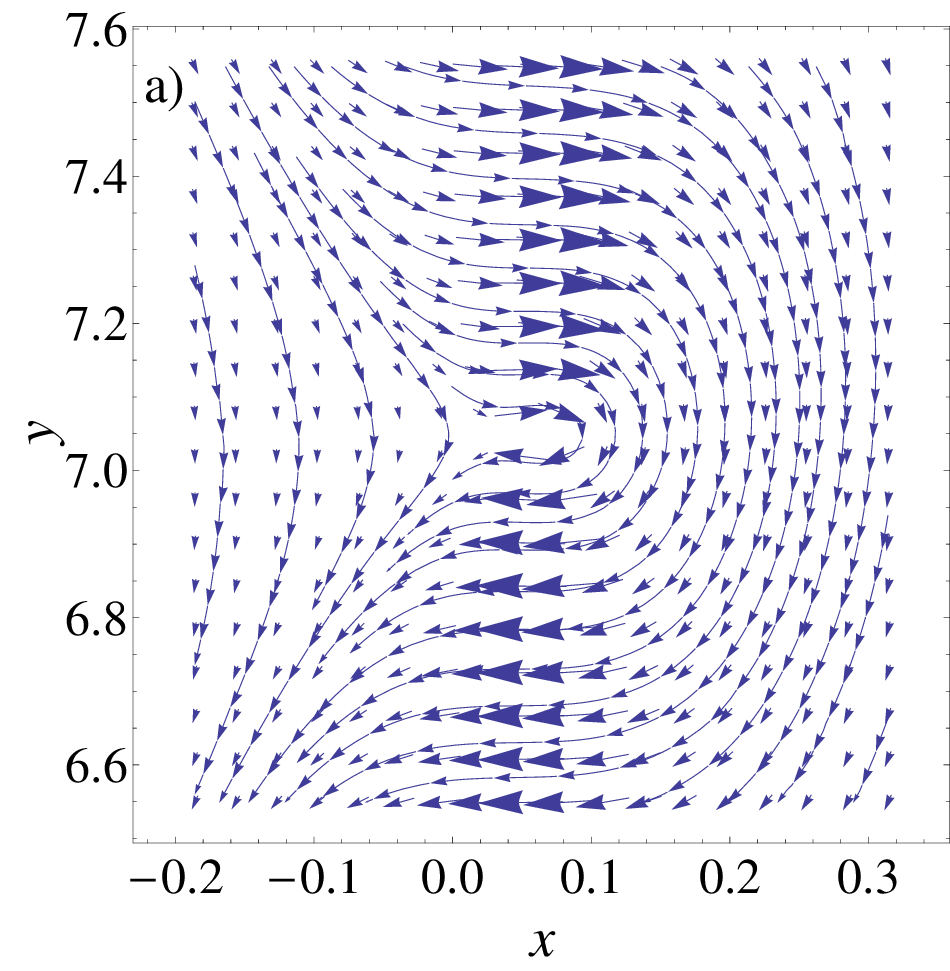}
\includegraphics[width=15.0pc]{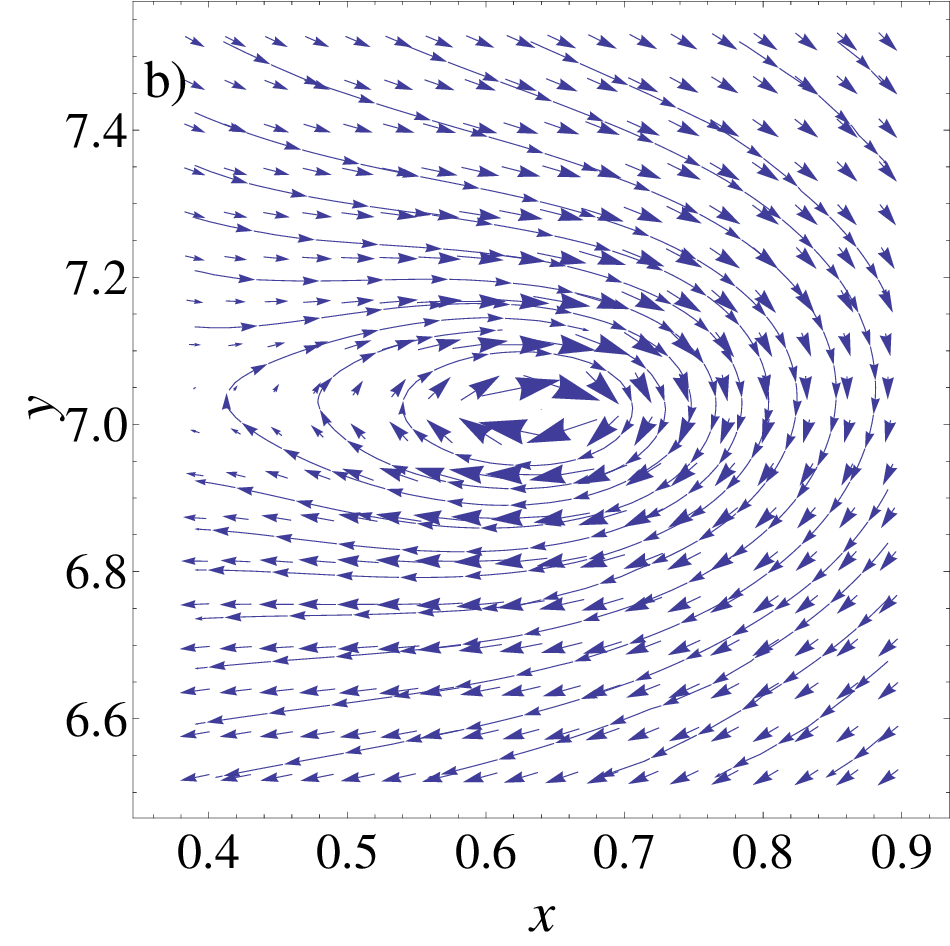}
\caption{\label{f4} Векторное поле $\mathbf{v}(x,y,t_{i})$ вблизи центров квантовых вихрей: a) $F_{0}=0.4$, $x_{0}=0.064$, $y_{0}=7.05$; b) $F_{0}=4$,  $x_{0}=0.64$, $y_{0}= 7.02$. Момент времени $t=5$.}
\end{figure*}

Из графиков видно, что увеличение напряжённости поля $F_{0}$ приводит к увеличению масштаба вихря. Построенные линии тока отчётливо демонстрируют, что в случае большой напряжённости $F_{0}$ поле скоростей $\mathbf{v}(x,y,t)$ имеет форму близкую к соленоидальной на гораздо больших масштабах, чем в случае малых напряжённостей (см. Рис.2 c),d)).

\section{Заключение}

В данной работе для фотоэлектрона, вырванного из двумерного атома водорода предельно коротким лазерным импульсом, была получена волновая функция, описывающая квантовые вихри. Аналитическое выражение этой волновой функции, как в импульсном, так и в координатном представлении имеет простой вид -- гауссова функция умноженная на полином второй степени относительно координат фотоэлектрона. Полином несет информацию о центрах квантовых вихрей и отвечает за вихревое поведение поля скоростей фотоэлектрона, а гауссова функция описывает расплывание образованного волнового пакета.

Полученная волновая функция позволила исследовать эволюцию квантового вихря в координатном пространстве: вихрь перемещается в пространстве без искажений со скоростью определяемой нулём волновой функции в импульсном представлении, соответствующим центру квантового вихря.

Показано, что масштаб вихря, формально определяемый как область в которой векторное поле скоростей имеет соленоидальную структуру, может быть изменён варьированием величины напряжённости электрического поля ионизирующего импульса.

\end{document}